\documentclass{aa}  

\usepackage{graphicx}
\usepackage{txfonts}
\usepackage{comment}
\usepackage[usenames]{xcolor}

\newcommand{\rr}[1]{{\sc #1}}

\newcommand\tabref[1]{%
Table~\ref{tab:#1}}

\newcommand\figref[1]{%
Fig.~\ref{fig:#1}}

\newcommand\secref[1]{%
Sec.~\ref{sec:#1}}

\newcommand{\Ms}{\,\mathrm{M_\odot}} 					
\newcommand{\keV}{\,\textrm{keV}}
\newcommand{\erg}{\,\textrm{erg}}
\newcommand{\Mpc}{\,\textrm{Mpc}}
\newcommand{\kpc}{\,\textrm{kpc}}
\newcommand{\cm}{\,\textrm{cm}}

\newcommand\ciao{\textsc{ciao}}
\newcommand\xspec{\textsc{Xspec}}
\newcommand{\Chandra}{{\it Chandra}}
\newcommand{\XMM}{{\it XMM-Newton}}
\newcommand{\Suzaku}{{\it Suzaku}}

\begin{document} 

\title{Constraints on the presence of a 3.5 keV dark matter emission line from Chandra observations of the Galactic centre}
\titlerunning{3.5 keV line emission constraints from the GC}
\author{Signe Riemer-S\o rensen\inst{1}\inst{2}}
\institute{
Institute of Theoretical Astrophysics, University of Oslo, PO 1029 Blindern, 0315 Oslo, Norway\\
\email{signe.riemer-sorensen@astro.uio.no}
\and
ARC Centre of Excellence for All-sky Astrophysics (CAASTRO)}
\date{\today}

 
  \abstract
   {Recent findings of line emission at $3.5\keV$ in both individual and stacked X-ray spectra of galaxy clusters have been speculated to have dark matter origin.}
   {If the origin is indeed dark matter, the emission line is expected to be detectable from the Milky Way dark matter halo.}
   {We perform a line search in public \Chandra{} X-ray observations of the region near Sgr A*. We derive upper limits on the line emission flux for the $2.0-9.0\keV$ energy interval and discuss their potential physical interpretations including various scenarios of decaying and annihilating dark matter.}
   {While find no clear evidence for its presence, the upper flux limits are not inconsistent with the recent detections for conservative mass profiles of the Milky Way.}
   {The results depends mildly on the spectral modelling and strongly on the choice of dark matter profile.}

   \maketitle

\section{Introduction}
Recent results by \citet{Bulbul:2014} and \citet{Boyarsky:2014} show a narrow excess of emission around $3.5\keV$ in both individual and stacked X-ray spectra of galaxy clusters. The spectral shape is consistent with mono-energetic line emission. There are no obvious astrophysical origins and alternative options such as dark matter have been considered \citep[][and references therein]{Bulbul:2014, Boyarsky:2014,Iakubovskyi:2015r}. If this is the case, the line should also be detectable from other dark matter dominated objects, in particular when stacking spectra from many different observations. Such dedicated line searches were performed in individual galaxy clusters \citep{Urban:2015,Iakubovskyi:2015}, individual galaxies \citep{Anderson:2014,Boyarsky:2014}, as well as stacked spectra of galaxies \citep{Anderson:2014} and dwarf galaxies \citep{Malyshev:2014}. With some searches providing low significance confirmation of the stacked galaxy cluster signal at $3.5\keV$, and some providing strong upper limits on the dark matter line emission flux \citep[][gives a detailed list of all searches]{Iakubovskyi:2015r}, the discussion is still ongoing. The nearest object that should provide detectable line emission if the origin is indeed dark matter, is the Milky Way halo providing a nearby consistency check. In this paper we analyse stacked X-ray spectra towards the Milky Way Centre to look for such a line but find no clear evidence for its existence. 

\section{Data} \label{sec:data}
We use a number of publicly available\footnote{\url{http://cda.harvard.edu/chaser/}} \Chandra{} X-ray observations of the region within 20\arcmin{} of Sgr A* located at (RA, Dec) = (17\,h\,45\,m\,40.0409\,s,  -29\degr \,0\arcmin \,28.118\arcsec) corresponding to galactic coordinates ($l,b$) = (359.944\degr, -0.04605\degr). The observation identification numbers (Obs ID) and exposure times are listed in \tabref{obs}. All the selected exposures are observed with the ACIS I0-I3 chips and consequently the total field of view is a square of $16.8$\arcmin$ \times$16.8\arcmin. The raw data were processed chip-by-chip using the software package, \ciao{} version 4.6 with {\sc caldb} 4.6.1,\footnote{\url{http://cxc.harvard.edu/ciao/}} following the ACIS data analysis guide.\footnote{\url{http://cxc.harvard.edu/ciao/guides/}} In brief the steps are:

{\bf Reprocessing:}  The files are reprocessed with \ciao{} to ensure the application of the most up-to-date calibration database and to set the observation specific bad pixel file so cosmic rays etc. are excluded from the analysis. All observations were taken in {\tt faint} mode, which uses a $3\times3$ pixel island to grade the events for bad pixel exclusion.

\begin{figure}
	\centering
	\includegraphics[width=0.99\columnwidth]{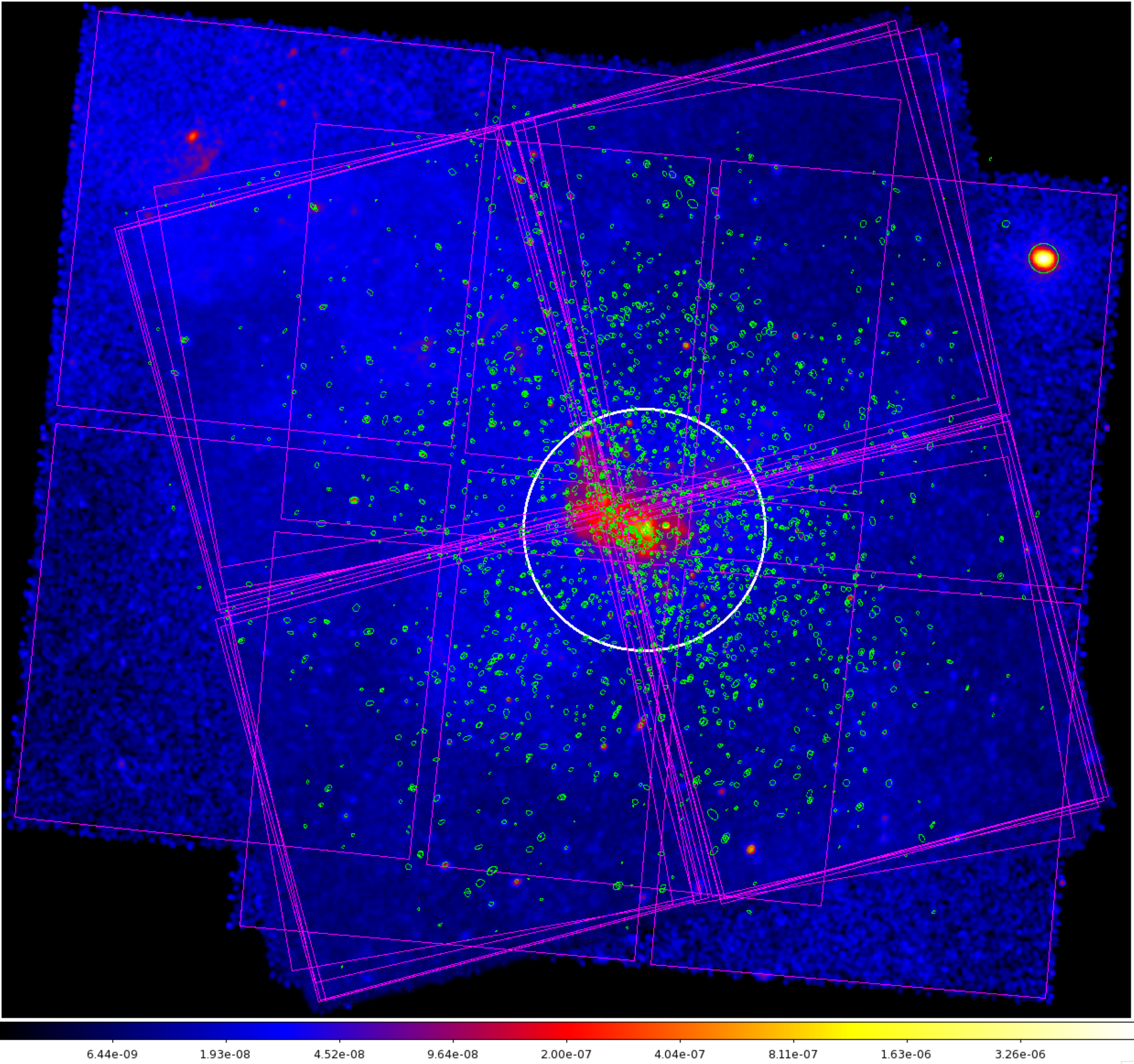}
	\caption{Fluxed image of the observations in \tabref{obs} showing the positions of each $8\arcmin\times8\arcmin$ chip for each observation (magenta squares) and the regions regions removed as point sources (small green ellipses, $<1\%$ of the total area) and Sgr A (white circle). The data have been smoothed with a 5 pixel Gaussian for visualisation purposes.} 
	\label{fig:regions}
\end{figure}

{\bf Source region selection:} In order to minimise calibration uncertainties we avoid the edges of each chip by cutting a square of 8\arcmin$\times$8\arcmin{} \citep{POG}. We also exclude a circle with radius of $2.5$\arcmin{} around Sgr A* as illustrated in \figref{regions}.

{\bf Point source removal and deflaring:} Point sources were removed from the images using the {\tt wavdetect}\footnote{\url{http://cxc.harvard.edu/ciao/ahelp/wavdetect.html}} routine in \ciao{} with the standard threshold value of $10^{-6}$ allowing for 1-2 spurious source detections per field. The detected sources are given in \figref{regions} where the ellipses have axes of 2-16\arcsec{} and a total area corresponding to less than 1\% of the chip area. Likewise we iteratively removed periods with flaring activity where the flux exceeded $\pm3\sigma$ of the mean using the {\tt lc\_sigma\_clipping}\footnote{\url{http://cxc.harvard.edu/ciao/ahelp/lc_sigma_clip.html}} routine. The light curves were visually inspected during the process and none of them contained strong flares extending over several bins. 

{\bf Background selection:} We do not remove any background from the observations. Rather we add the line emission visible in observations taken with the telescope in the stowed position (``lid on'')\footnote{\url{http://cxc.cfa.harvard.edu/contrib/maxim/stowed/}} to the model later. These are denoted ``instrumental lines'' in \tabref{lines} \citep{POG,Bartalucci:2014}. \figref{regions} clearly shows pockets of extended diffuse emission, on scales smaller than expected for dark matter emission from the halo. While exclusion of these regions would improve the expected dark matter signal to noise, they would also decrease the observed field of view significantly and with the danger of removing underlying halo substructure.

{\bf Spectrum extraction:} The spectra were extracted for the source regions described above using {\tt specextract}\footnote{\url{http://cxc.harvard.edu/ciao/ahelp/specextract.html}} with the default settings apart from the rebinning, which was set to at least 15 counts per bin. The instrumental response files necessary to perform the analysis in physical units and to compensate for a non-uniform effective collecting area on the detector were also computed: the redistribution matrix files (RMFs) and the ancillary response files (ARFs).

\begin{figure*}
	\centering
	\includegraphics[trim = 30mm 10mm 15mm 18mm, clip,width=0.99\columnwidth]{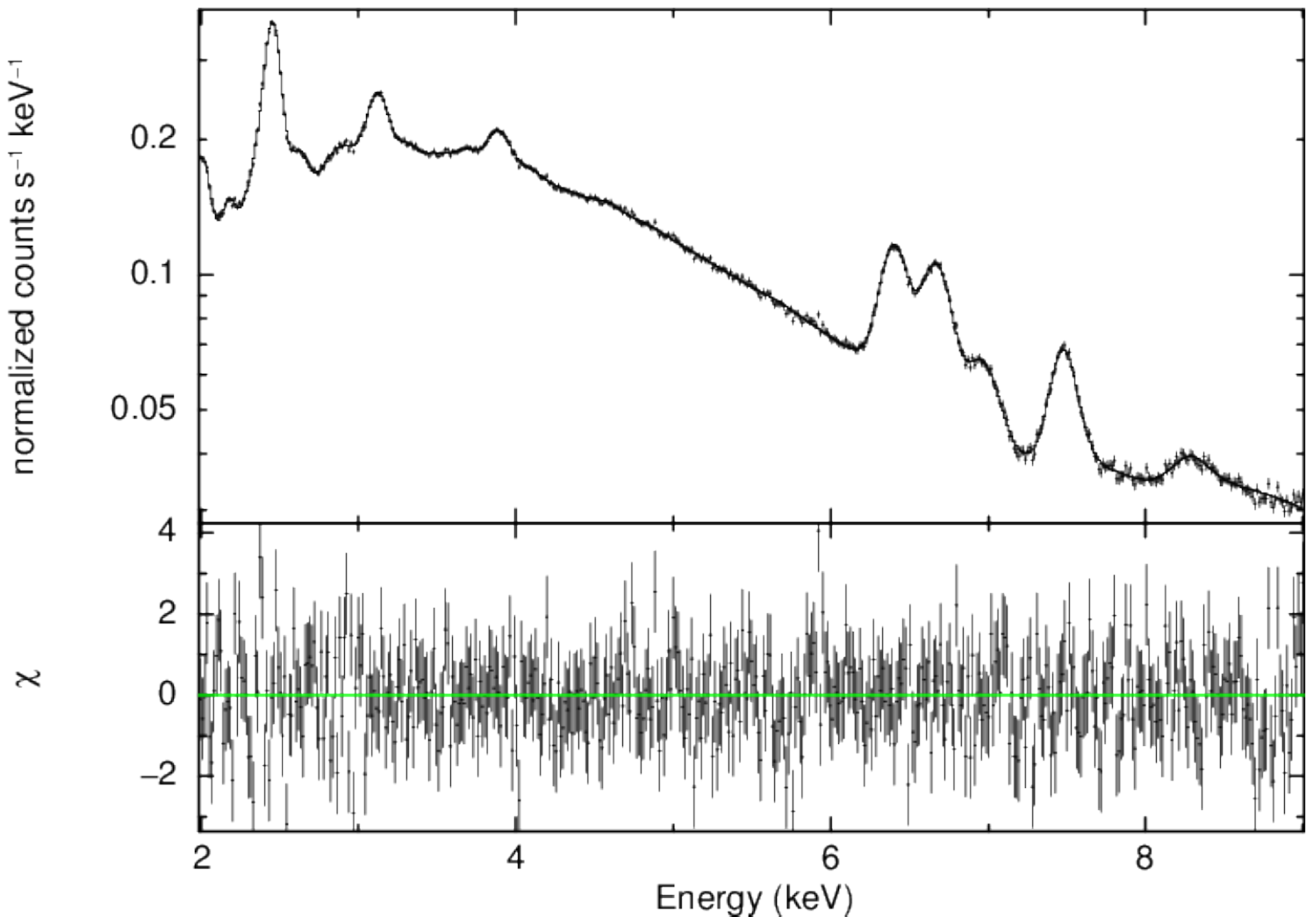}
	\includegraphics[trim = 30mm 10mm 15mm 18mm, clip,width=0.99\columnwidth]{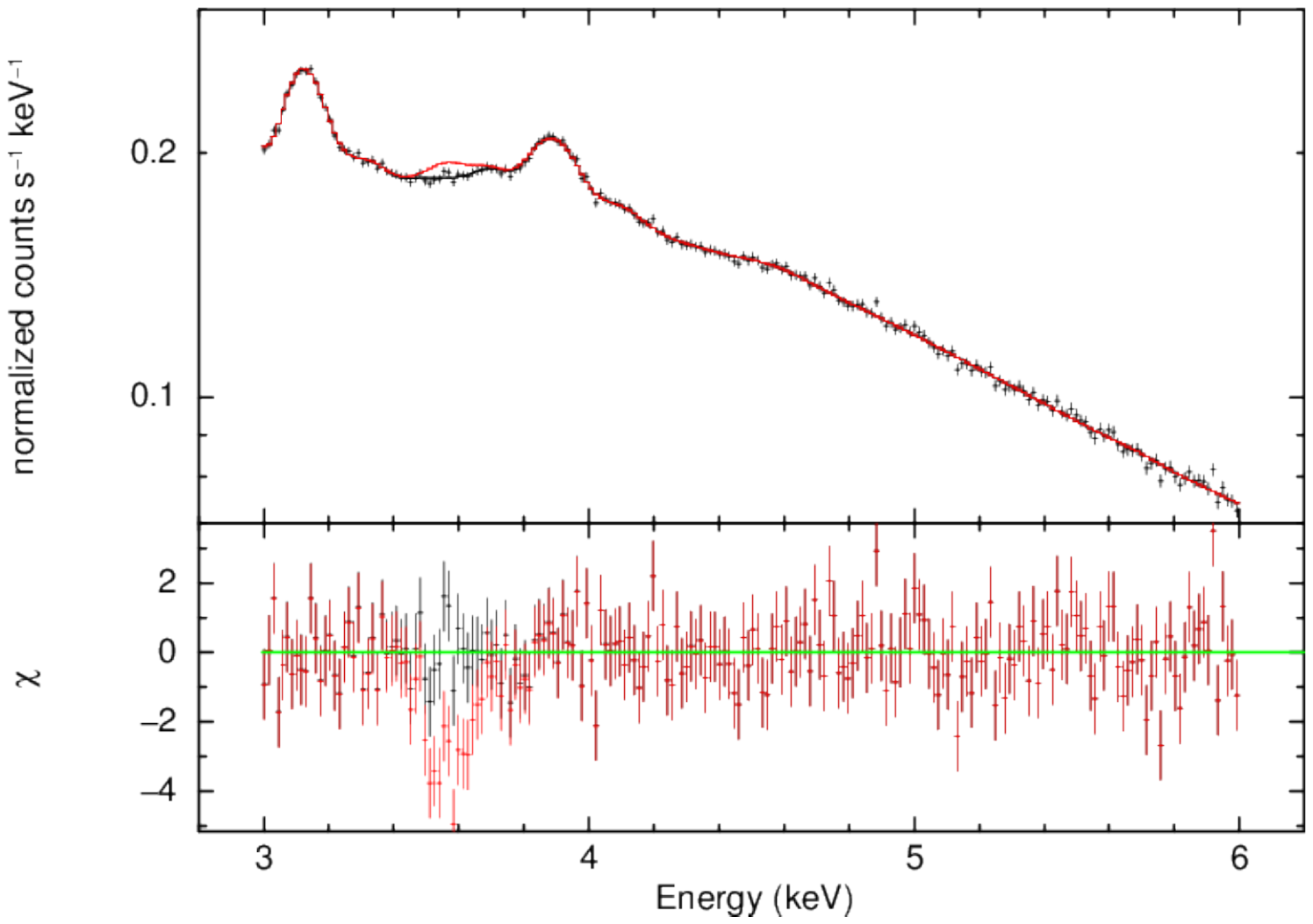}
	\caption{{\bf Left:} The stacked spectrum and best-fit base model (without any additional line emission) in normalised counts $\keV^{-1}\sec^{-1}$, and $\chi$ residuals below for the $2.0-9.0\keV$. No significant emission line excess is seen around $3.5\keV$. 
	{\bf Right:} Same for the $3.0-6.0\keV$ interval. The red curve shows the model including the mass-scaled expected signal from \citet{Bulbul:2014} and residuals under the assumption of an NFW profile for the Milky Way dark matter halo.}
	\label{fig:stacked}
\end{figure*}

{\bf Stacking:} The spectra and response files are weighted and stacked using the {\tt combine\_spectra} routine in \ciao{} providing a summed source spectrum, an exposure-weighted source ARF, a source RMF weighted by exposure time and ARF, and an area- and exposure-weighted background spectrum. The resulting stacked spectrum of all observations is shown in \figref{stacked}.

\begin{table}
\caption{\label{tab:obs} Public \Chandra{} observations used in the analysis.}
\begin{tabular}{ l l l l l l}
\hline \hline
Obs  	& Exposure 	& Cleaned		& R.A.	& Dec.	& Angle  \\
ID		& [ks] 		& [ks]  		& [deg]	& [deg]	& [deg]  \\ \hline

3392		& 167.0 			& 166.0	& 266.419	& -29.004	& 75.5 \\
3393		& 160.1 			& 157.7 	& 266.420	& -29.004	& 75.5 \\
3665		& 91.1 			& 89.7	& 266.420	& -29.004	& 75.5 \\
5953		& 46.0 			& 45.1 	& 266.415	& -29.012	& 275.3\\
10556	& 114.0 			& 112.2 	& 266.416	& -29.000	& 79.0 \\
11843	& 80.0			& 78.8 	& 266.415	& -29.000 & 80.7\\
13438	& 67.0			& 66.	0	& 266.502	& -28.979 & 275.8\\
 \hline
Total		& 825.0 			& 750.6 \\ \hline
\end{tabular}
\end{table}

\section{Spectrum modelling} \label{sec:model}
Modelling the astrophysical X-ray emission of the Milky Way centre is notoriously difficult because many sources contribute with a range of different signatures \citep[e.~g.][]{Revnivtsev:2007} that may hide the line we are looking for. However, since we are not interested in the physical properties of the unresolved sources contributing to the broad features of the background emission, we do not require the model to represent the underlying physics, allowing us to chose the simplest possible model that provides a good fit to the data.

\subsection{Energy intervals}
We use the \xspec{} 12.8 fitting package for analysing the spectra.\footnote{\url{http://heasarc.gsfc.nasa.gov/xanadu/xspec/}} We consider two energy intervals for the base model fit namely the broad interval of $2.0-9.0\keV$, and the narrower interval of $3.0-6.0\keV$. Below $2\keV$ the spectrum is dominated by astrophysical emission lines that cannot be individually resolved. Above $9\keV$ there is a wide complex of instrumental Au lines. Similarly to \citet{Bulbul:2014} we examine the $3-6\keV$ interval thoroughly due to the possible excess around $3.5\keV$. This interval is chosen to be wide enough to measure the continuum accurately and avoiding the strong S and Si lines below $3\keV$ and Fe lines above $6 \keV$.

\subsection{Base model}
The base for our explorations is a line-free apec model \citep[i.e. with the abundance parameter fixed to zero][]{Smith:2001} combined with a power law and eight Gaussians to mimic residual detector background \citep{Bartalucci:2014} and a photo-electric absorption model (wabs) to account for galactic absorption. Following \citet{Muno:2004} we assume a fixed absorption with an average equivalent hydrogen column density of $6\times10^{22}\, \mathrm{atoms} \cm^{-2}$. If we fit the $2.0-9.0\keV$ interval for the absorption we obtain a value of $5.83_{-0.029}^{+0.037}\times 10^{22}\mathrm{atoms} \cm^{-2}$, consistent with the fixed value. The line-free model is pre-fitted to intervals that appear line-free ($3.4-3.6\keV$ and $4.3-5.2\keV$), before we add the known instrumental lines from \citet{Bartalucci:2014} and atomic emission lines (all listed in \tabref{lines}).

The best fit normalisations of the instrumental lines are of the same order of magnitude as those given in the literature for blank sky spectra \citep{Bartalucci:2014,POG}. 

For the atomic lines we include all lines with emissivities larger than $5\times10^{-19} \mathrm{photons\,} \mathrm{cm}^{-3} \sec^{-1}$ for a plasma temperature of $2\keV$, as listed in \tabref{lines} (this corresponds to the atomic lines used in \citet{Bulbul:2014}). The central values of the lines are allowed to ``budge'' by $0.01\keV$ in the fit to account for uncertainties and mis-calibrations in the spectra. The widths are assumed to be unresolved, but it was checked that allowing for broadening up to $0.1\keV$ does not improve the fit quality. The normalisations are free to vary and we do not make any assumptions about their internal relations. Atomic lines that do not change the best fit $\chi^2$ are removed from the final model (marked in \tabref{lines}). 

Since atomic emissivities depend on the plasma temperature, which is more complicated towards the centre of the Milky Way than in galaxy clusters, the list in \citet{Bulbul:2014} may not be sufficient here. To account for this as well as possible instrumental artefacts, a number of extra lines are added to the fit with the central energies and normalisations as free parameters while fitting the base model. They do not correspond to any known atomic transitions but are simply a parameterisation of our ignorance. Only the broad interval fit is improved by such additional lines at the best fit energies of $4.84\keV$ and $9.01\keV$ (listed in \tabref{lines}). The origins of the additional lines are unknown and could potentially be due to non-astrophysical processes such as decaying dark matter. The normalisations of the lines are similar to the upper limits on line emission at other energies, and we leave the interpretation of the lines to the reader, but note that the upper flux limits presented in \secref{general} and \ref{sec:sterile} remain valid even at those energies because we do not remove their contribution from the flux constraints but rather include them in the total flux.

\begin{table*}
    \caption{\label{tab:basemodel} Line free base model and instrumental lines.}
    \begin{tabular}{lll | ll}
    \hline \hline
\multicolumn{3}{l |}{Line-free model}			&												\multicolumn{2}{l}{Instrumental lines}  \\ 
Parameter								& Best fit broad					& Best fit narrow		& Energy [keV]						& Width [keV] \\ \hline 	
wabs nH $[\cm^{-2}]$ (frozen)				& $6\times10^{22}$ 				& $6\times10^{22}$		& 1.12\tablefootmark{a}				& 1.36 		\\ 		
apec kT $[\keV]$						& $2.15^{+1.29}_{-0.01}$			& $2.14^{+0.92}_{-0.57}$	& 1.49\tablefootmark{a} (Al K-$\alpha$)	& 0.492		\\		
apec Abundance [solar] (frozen)			& $0$ 	 					& $0$ 				& 1.81\tablefootmark{a} (Si K-$\alpha$)	& 0.492 		\\		
apec redshift (frozen)					& $0$						& $0$ 				& 2.15\tablefootmark{a} (Au M-$\alpha\beta$) & unresolved \\ 		
apec norm $[10^{-2} \cm^{-5}]$				& $3.28^{+1.88}_{-0.12}$			& $2.18^{+0.95}_{-0.05}$	& 5.90\tablefootmark{a}				& 0.636		\\		
power law PhoIndex						& $1.27^{+0.08}_{-3.12}$			& $1.16^{+0.47}_{-0.61}$	& 7.48\tablefootmark{a}				& unresolved	\\		
power law norm	$[10^{-3}\keV^{-1}\cm^{-2}\sec^{-1}]$	& $2.78^{+7.1}_{-0.01}$		& $2.71^{+3.35}_{-0.05}$	& 8.31\tablefootmark{a}				& 0.071 		\\		
									&							&					& 9.71 (Au L-$\alpha$)				& 1.23		\\ \hline
\end{tabular}
\tablefoot{
The large uncertainties on the apec and power law models are due to degeneracy between the two models. However, even the extreme case of removing one or the other from the base model leads to very similar flux constraints, and we take the sum of the two to be robustly determined.\\
\tablefoottext{a}{Lines removed in the narrow interval model.}
}
\end{table*}

\begin{table*}
\caption{\label{tab:lines} Emission lines included in the base model.}
    \begin{tabular}{ll  ll  ll  |  lll}
    \hline \hline
\multicolumn{6}{l |}{Emission lines}																									& \multicolumn{3}{l}{Additional lines} \\				
Energy  					& Line  		& Energy 							& Line  		& Energy						& Line  			& Energy 			& Width	& Normalisation \\ 
$[\keV]$ 					&	  		& [keV] 							& 	  		& [keV] 						& 	  			& [keV]			& [keV]	& [photons$\cm^{-2}\sec^{-1}$] 			\\ \hline
2.01						& Si \rr{xiv}	& 3.47							& K \rr{xviii}	& 6.39\tablefootmark{a}			& Fe K-$\alpha$ 	& 4.84\tablefootmark{a}			& unresolved & $1.04\times10^{-6}$\\
2.05\tablefootmark{a}		& Al \rr{xiii}	& 3.51							& K \rr{xviii} 	& 6.70\tablefootmark{a}			& Fe \rr{xxv}		& 9.01\tablefootmark{a}			& unresolved & $5.88\times10^{-6}$\\
2.18\tablefootmark{a}		& Si \rr{xii}		& 3.62							& Ar \rr{xvii}	& 6.62\tablefootmark{a}			& Fe \rr{xxiv}		& &\\
2.29\tablefootmark{a}		& Si \rr{xii}		& 3.68							& Ar \rr{xvii}	& 6.95\tablefootmark{a}			& Fe \rr{xxiv}		& &\\
2.34\tablefootmark{a}		& Si \rr{xii}		& 3.71							& K \rr{xix}	& 7.29\tablefootmark{a}\tablefootmark{b}& Fe \rr{xxv}	& &\\
2.45\tablefootmark{a}		& Si \rr{xv}	& 3.86							& Ca \rr{xix}	& 7.79\tablefootmark{a}			& Ni \rr{xxvii}		& &\\
2.51\tablefootmark{a}		& Si \rr{xiv}	& 3.90							& Ca \rr{xix}	& 7.81\tablefootmark{a}			& Fe \rr{xxv}		& &\\
2.62\tablefootmark{a}		& Si \rr{xiv}	& 3.93							& Ar \rr{xviii}	& 7.88\tablefootmark{a}			& Fe \rr{xxv} 		& &\\
2.88						& Si \rr{xv} 	& 4.10							& Ca \rr{xx}	& 8.29\tablefootmark{a}			& Fe \rr{xxv} 		& &\\
3.12						& Ar \rr{xvii} 	& 4.58							& Ca \rr{xix}	& 8.30\tablefootmark{a}			& Fe \rr{xxvi}		& &\\
3.31						& Ar \rr{xviii} 	& 5.69\tablefootmark{a}\tablefootmark{b}	& Cr \rr{xxiii}	& 8.70\tablefootmark{a}\tablefootmark{b}& Fe \rr{xxvi}	& &\\ \hline
\end{tabular}
\tablefoot{Known astrophysical emission lines included in the model together with additional unknown unresolved lines. Notice none of the additional lines lies in the vicinity of $3.5\keV$.
\tablefoottext{a}{Removed in narrow interval model due to no influence on $\chi^2$.}
\tablefoottext{b}{Removed in broad interval model.}
}
\end{table*}

\begin{figure*}
	\centering
	\includegraphics[trim = 40mm 18mm 25mm 18mm, clip,width=0.99\columnwidth]{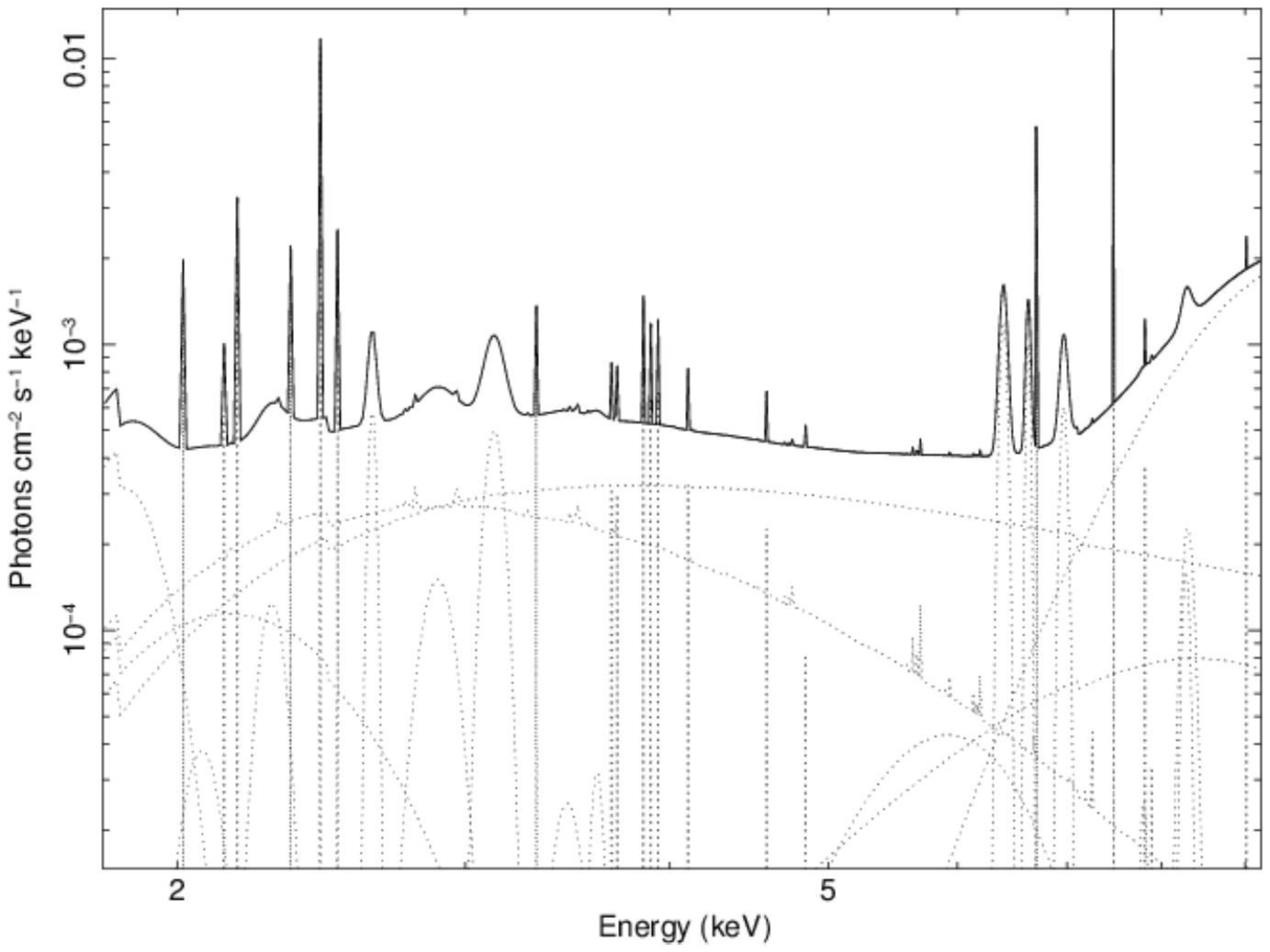}
	\includegraphics[trim = 40mm 18mm 25mm 18mm, clip,width=0.99\columnwidth]{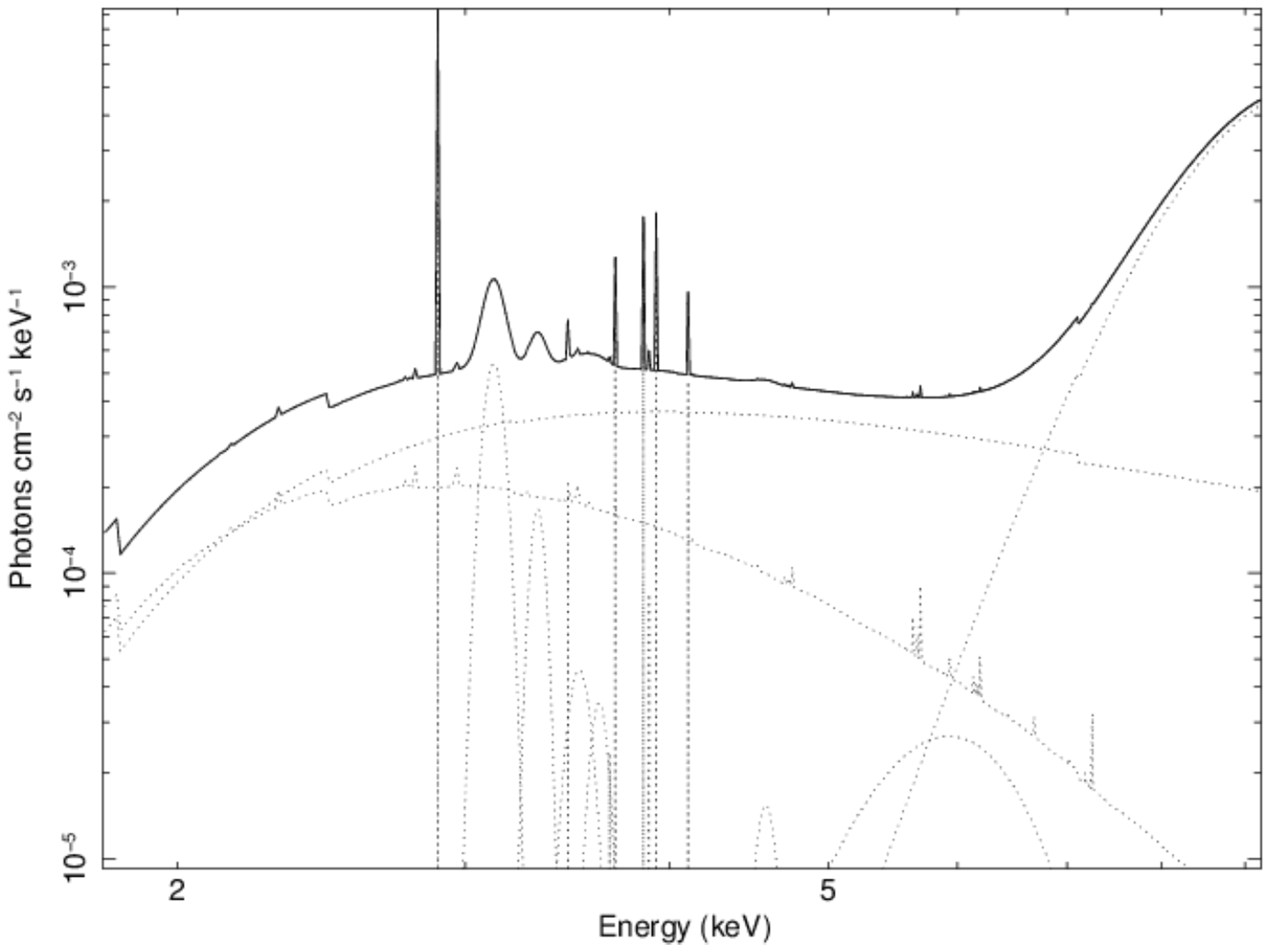}
	\caption{{\bf Left:} The best fit incident model spectrum of the base model showing the contributions of the individual additive model components for the $2.0-9.0\keV$ interval. {\bf Right:} Same for the $3.0-6.0\keV$ interval.}
	\label{fig:model}
\end{figure*}

The final base model with best fit with $\chi^2$/dof$=621.9/406=1.53$ and $172.02/172 = 1.00$ for the broad and narrow intervals respectively is shown in \figref{model}. The fit quality is somewhat worse for the broad interval due to the large number of S and Si lines below $3\keV$ and the Fe lines at $6-8\keV$ despite our attempt to include all the lines in the model. The residuals in \figref{model} are random, with no significant structures at the scale of the instrumental resolution. Where overlapping, the results from the two intervals are consistent.

\subsection{Additional line}
On top of the best fit base model we add an unresolved Gaussian to account for non-astrophysical line emission. The galactic absorption is also applied to the additional line so the entire model consist of {\it absorption $\times$ (continuum + atomic and extra lines + possible dark matter emission)}. For each central energy in $0.05\keV$ steps, we fit the model to the stacked spectrum, and then increase the normalisation of the additional Gaussian and refit all free parameters until $\Delta\chi^2 = \chi^2(\mathrm{norm})-\chi^2(\mathrm{best\, fit}) = 4$ corresponding to the upper 95\% confidence level for one degree of freedom (the normalisation) marginalised over all other parameters.\footnote{Following the consensus in the literature this is the two-sided limit even though a one-sided limit $\Delta\chi^2 = 2.71$ would be more meaningful.} The line flux is calculated in bins of Full Width Half Max (FWHM) of the spectral resolution around the central value approximated by \citep{POG}
\begin{equation}
\Delta E_{FWHM}=0.012 E_\gamma + 0.12 \keV \, .
\end{equation}

For each energy we regard the \emph{entire flux in all Gaussians} including atomic lines, instrumental lines, and the additional lines as an upper limit on the possible emission from dark matter. Only the smooth component is subtracted from the model, resulting in a very conservative limit as the line flux can have astrophysical origin, but the method takes into account the risk of a non-astrophysical line hiding under an astrophysical emission line. We also present the results of subtracting the instrumental features with widths much wider than the instrumental resolution (1.12, 1.49, 1.81, 9.71 keV). This mainly affects energies above 4 keV.

\begin{figure}
	\centering
	\includegraphics[width=0.99\columnwidth]{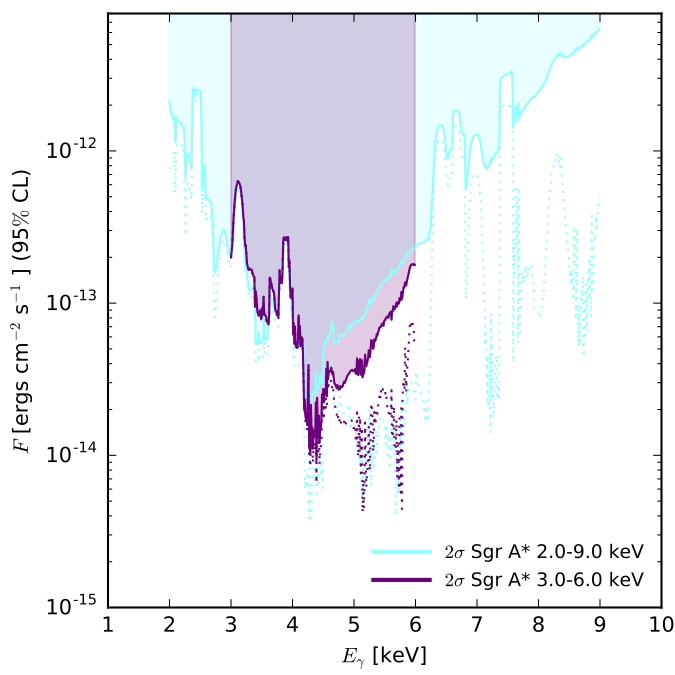}
	\caption{{\bf Left}: Constraints on line emission flux in the stacked spectrum (95\% confidence). The cyan lines are for the entire 2-9\keV{} interval model, and the purple lines are for the 3-6\keV{} interval. The solid lines represents the total flux in line emission (including instrumental and astrophysical lines while for the dotted lines, the broad instrumental lines have been subtracted.} 	
	\label{fig:flux}
\end{figure}

\citet{Bulbul:2014} estimate the maximum contribution from atomic lines around $3.5\keV$ by using the ratios of different species. Because of the dependence on the assumed plasma conditions \citep[e.g. temperature,][]{Jeltema:2014,Jeltema:2014_reply}, the method cannot easily be transferred to the Milky Way, and is not applied here. The resulting flux limits are shown in \figref{flux}. 

\section{Mass within field of view} \label{sec:mass}
The dark matter emission scales with the amount of dark matter within the observed field of view. For decaying dark matter the emission is directly proportional to density, whereas for annihilating dark matter it scales with density squared. We calculate the mass within the field of view as the integral along the line of sight of a $16$\arcmin$\times16$\arcmin{} square, assuming the density variation over this region to be negligible\footnote{Integrating along the line of sight over patches with radius of $0.5\arcsec$ (corresponding to the resolution of \Chandra{}) the variation across the field of view (including the two exposures that are slightly offset from the Galactic centre) is $<10\%$.}. We subtract the area of a circle with radius of $2.5$\arcmin{} to account for the removed region around Sgr A* (7\%, taking the gaps between the chips into account changes this by $<1\%$) and an additional 1\% to account for the removed point sources. The total area of the field of view becomes 16\arcmin$\times16$\arcmin{}$\times0.92-2\pi(2.5\arcmin)^2 = 196.25$\arcmin$^2$. The uncertainty on the final mass within the field of view arising from the choice of mass profile is much larger than any uncertainty on the area being covered by the observations.

For the dark matter density profile as a function of radius ($r$), we consider various profiles. The Einasto profile \citep{Einasto:1989} is based on observed stellar density profiles,
\begin{equation}
\rho_\mathrm{DM}(r) = \rho_\odot\exp{\left[ -\frac{2}{\alpha} \left(  \left(\frac{r}{r_\mathrm{s}} \right)^\alpha   -\left(\frac{r_\odot}{r_\mathrm{s}} \right)^\alpha \right)  \right]} \, ,
\end{equation}
with $r_\mathrm{s}$ being the scale radius, $\rho_\odot$ the density at solar radius $r_\odot$ from the galactic centre, and $\alpha$ is a constant. The parameter values  and the results of integrating over density and density squared are given in \tabref{mfov}. The main uncertainty is on the $\alpha$ parameter, which simulations show can vary quite significantly \citep{Dutton:2014}. We adopt a default value of $\alpha=0.17$ \citep{Dutton:2014} as well as the larger value of $0.20$ \citep{Tissera:2010} representing a more cored profile and leading to $\sim 40\%$ uncertainty on the density integral and a factor of two for the squared integral.

Simulations indicate that the dark matter profiles may be steeper than the Einasto profile for cold dark matter \cite[Navarro-Frenk-White, NFW]{Navarro:1997} and flatter for warm dark matter \cite[isothermal]{Weber:2010} or when including baryons \citep{Dutton:2014}. Both cases can be described by the generalised profile
\begin{equation}
\rho_\mathrm{DM}(r) = \rho_\odot \frac{ \left( \frac{r_\odot}{r_\mathrm{s}}\right)^\gamma  \left(1+\left( \frac{r_\odot}{r_\mathrm{s}} \right)^\alpha \right)^{(\gamma-\beta)/\alpha} }{\left( \frac{r}{r_\mathrm{s}}\right)^\gamma  \left(1+\left( \frac{r}{r_\mathrm{s}} \right)^\alpha \right)^{(\gamma-\beta)/\alpha}} \, ,
\end{equation}
with $\alpha=\gamma=1$ and $\beta=3$ for the NFW profile, and $\alpha=\beta=2$ and $\gamma=0$ for an isothermal profile. 

For the NFW profile, the observed parameter range given in \tabref{mfov} leads to a factor of two uncertainty on the density integral and an order of magnitude on the density squared. For the isothermal profile the integrated mass is smaller than the NFW profile by a factor of 20 when using the parameters from \citet{Sofue:2010er}, but their value of the local dark matter density is somewhat smaller than recent dynamical constraints \citep{Pato:2015}, which we will adopt here. The squared integral is smaller by three orders of magnitude.

The differences between the various profiles are much larger than the uncertainties due to parameter uncertainties. For the sterile neutrinos that will be discussed in \secref{sterile}, their behaviour and expected dark matter profiles depend on the production mechanism, but resonant production leads to cold dark matter like structures. Consequently, we adopt the NFW profile which also lies in the middle of the range of values in \tabref{mfov}. We also show the effect of using the isothermal profile with the same local dark matter density \citep{Sofue:2010er,Pato:2015}.
The uncertainty of mass profile selection can be avoided if one look at the outskirts of dark matter halos rather than the centers, but the expected signal will also be weaker.

\begin{table*}
\caption{\label{tab:mfov} Density profiles and masses within field of view.}

\begin{tabular}{l l l l l l l l}
\hline \hline
Profile 								& $\rho_\odot $					& $r_\odot$ & $r_\mathrm{s} $ 		& $\int \rho(r) \mathrm{d} V$			& $D_\mathrm{avg}$	 	& $\int \rho^2(r) \mathrm{d} V$				& $D_\mathrm{avg}$ \\ 
									& $[\Ms \kpc^{-3}]$				& $[\kpc]$ & $[\kpc]$ 			& $[M_\odot]$						& $[\kpc]$	 			& $[M_\odot^{2} \kpc^{-3}]$				& $[\kpc]$ \\ \hline
Einasto\tablefootmark{1} (a) $\alpha=0.17$	& $10.5^{+0.8}_{-0.8} \times 10^6$	& 8.3		& $20$				& $3.0^{+0.2}_{-0.2}\times10^6 $		& 8.8					& $6.2^{+0.9}_{-1.0}\times10^{15}$			& $8.3$ \\
Einasto\tablefootmark{2} $\alpha=0.20$		& $10.5^{+0.8}_{-0.8} \times 10^6$	& 8.3		& $20$				& $2.3^{+1.7}_{-1.7}\times10^6$		& 8.9					& $2.2^{+0.3}_{-0.3}\times10^{15}$			& $8.3$ \\
NFW\tablefootmark{3}					& $4.7^{+0.6}_{-0.6} \times 10^6$	& 8.0		& $21.0^{+3.2}_{-3.2}$	& $1.1^{+0.2}_{-0.2}\times10^6$		& $8.9$ 				& $1.2^{+0.7}_{-0.4}\times10^{15}$			& $8.3$ \\
NFW\tablefootmark{4} (b)					& $10.5^{+0.8}_{-0.8} \times 10^6$	& 8.3		& $19^{+7.5}_{-5.5}$		& $2.7^{+1.2}_{-0.5}\times10^6$		& $8.9$ 				& $7.4^{+9.8}_{-3.0}\times10^{15}$			& $8.3$ \\
NFW\tablefootmark{5}					& $14.0^{+10.0}_{-10.0} \times 10^6$& 8.1	& $16.1^{+8}_{-8}$		& $3.8^{+6.7}_{-2.9}\times10^6$		& $8.7$ 				& $15.3^{+124}_{-14.5}\times10^{15}$		& $8.3$ \\
Isothermal\tablefootmark{6} (c)				& $3.5^{+0.4}_{-0.4} \times 10^6$	& 8.0		& $12.0^{+1.3}_{-1.3}$	& $0.13^{+0.02}_{-0.01}\times10^6$		& $15.2$ 				& $4.5^{+1.5}_{-1.1}\times10^{11}$			& $11.1$ \\
Isothermal scaled\tablefootmark{7} (d)		& $10.9^{+0.8}_{-0.8} \times 10^6$	& 8.0		& $12.0^{+1.3}_{-1.3}$	& $0.39^{+0.03}_{-0.03}\times10^6$		& $15.2$ 				& $4.1^{+1.0}_{-0.7}\times10^{12}$			& $11.1$ \\ \hline
\end{tabular}
\tablefoot{Mass within field of view ($16$\arcmin$\times16$\arcmin{}$\times0.92\mathrm{\,(point\ sources\ and\ edges)}-2\pi(2.5\arcmin)^2 \mathrm{\,(Sgr A*)} = 196.25$\arcmin$^2$) and mass-weighted average distances for different dark matter profiles of the Milky Way. The (a, b, c, d) refer to corresponding labels on \figref{mass}. The parameter values are taken from: 
\tablefoottext{1}{\citet{Dutton:2014,Bernal:2012,Pato:2015},} 
\tablefoottext{2}{\citet{Tissera:2010,Bernal:2012,Pato:2015},} 
\tablefoottext{3}{\citet{Xue:2008},} 
\tablefoottext{4}{\citet{Bernal:2012,Read:2014,Pato:2015},} 
\tablefoottext{5}{\citet{Nesti:2013},} 
\tablefoottext{6}{\citet{Sofue:2009,Sofue:2010er},} 
\tablefoottext{7}{\citet{Sofue:2010er,Pato:2015}.}
}
\end{table*}

\begin{figure}
	\centering
	\includegraphics[width=0.99\columnwidth]{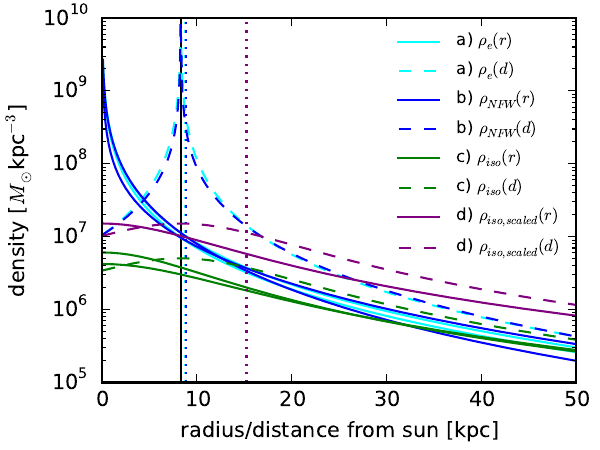}
	\caption{A selection of density profiles and parameter ranges from \tabref{mfov} as a function of radius from the galactic centre (solid) and along the line of sight towards Sgr A* (dashed). The vertical black line indicates the solar radius, and the dotted vertical lines are the mass-weighted average distances for each of the profiles.} 	
	\label{fig:mass}
\end{figure}

\section{General constraints} \label{sec:general}
In \figref{flux} (left) we present the general flux constraints on line emission near Sgr A*. In the shaded regions line emission is excluded at 95\% confidence based on all line emission in the model. 
These constraints apply to all dark matter candidates with mono-energetic photon emission proportional to the density whether from decays or other processes. For decays, the probability is proportional to density while for annihilation-like processes the interaction requires two particles and the probability is proportional to density squared. This includes scattering processes.

\subsection{Decay like dark matter}
The allowed decay rate leading to photon emission is shown in \figref{Esigmav} (left) for one photon per decay. For two-photon interactions, the constraints are a factor of two stronger. There is also a factor of two depending on whether the particles are of Majorana or Dirac type (because the decay probability is linear in number density). Here we assume the particles to be of Majorana type (their own anti-particles). If they instead are Dirac particles, the constraints are weakened by a factor of two.

These constraints are relevant to e.g. models where the dark matter decays to axions that are then converted to photons in magnetic fields \citep[e.g.][]{Cicoli:2014, Conlon:2014}. This can explain a signal from galaxy clusters, which would be absent in the Milky Way because the axion-photon conversion depends on the magnetic field strength \citep[e.~g.][]{Conlon:2014,Alvarez:2015}.

\subsection{Annihilation like dark matter}
For annihilating dark matter, the local flux is given by

\begin{eqnarray}
F_\mathrm{obs} &=& \frac{L_\gamma}{4\pi D_L^2} \\ \nonumber 
 &=& \frac{1}{4\pi D_L^2} \frac{N_\gamma E_\gamma}{2m^2} \int \sigma v(r)\rho^2(r) \mathrm{d} V \, ,
\end{eqnarray}

where $m$ is the particle mass, $N_\gamma$ is the number of photons per interaction, $E_\gamma$ is the photon energy, $\sigma$ is the interaction cross section, $v(r)$ is the velocity distribution of the dark matter as a function of radius and the integral is over the observed volume. For a back-of-the-envelope calculation we assume two photons with $E_\gamma = m$ per annihilation and that $\left<\sigma v \right>$ is independent of radius. Reordering and converting units we get


\begin{eqnarray}
\left<\sigma v \right> [\cm^{-3}\sec^{-1}] &=& 1.74\times10^{-11} \cm^{-3}\sec^{-1}\\ \nonumber
&\times& F_\mathrm{obs}[\erg\cm^{-2}\sec^{-1}] \\ \nonumber
&\times& \frac{D_L^2[\kpc^2]}{\int \rho^2(r) \mathrm{d} V[M_\odot^2\kpc^{-3}]}  \frac{m^2[\keV^2]}{E_\gamma[\keV]}\, .
\end{eqnarray}

\begin{figure*}
	\centering
	\includegraphics[width=0.99\columnwidth]{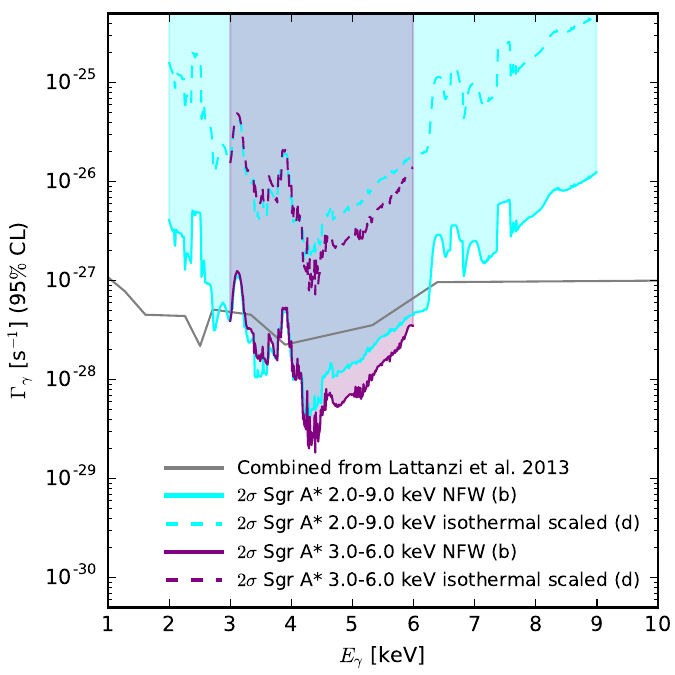}
	\includegraphics[width=0.99\columnwidth]{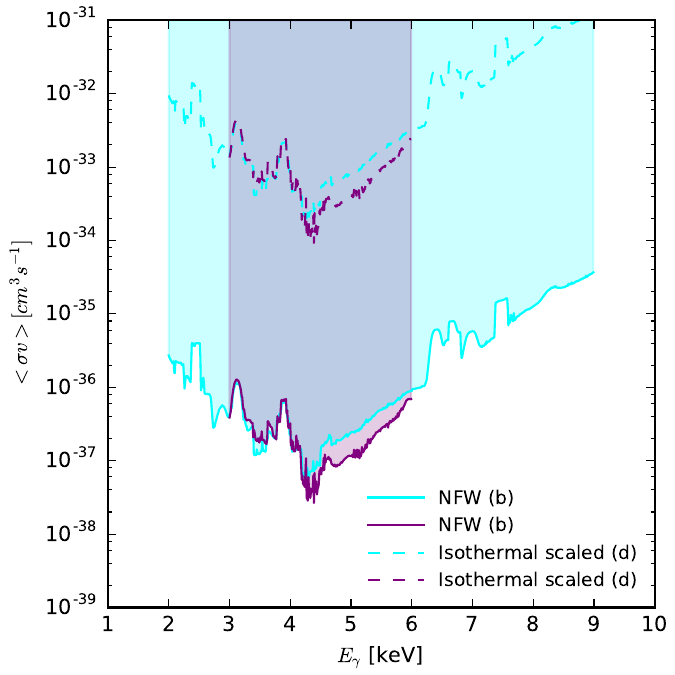}
	\caption{{\bf Left:} General constraints on the interaction rate with one-photon emission proportional to matter density (95\% confidence) for an isothermal (dashed) and NFW profile (solid). The grey lines show the combination of previous constraints \cite[][and references therein]{Lattanzi:2013}. The cyan lines are for the entire 2-9\keV{} interval model, and the purple lines are for the 3-6\keV{} interval.
	{\bf Right:} Constraints on the velocity averaged cross-section for models with two-photon emission from dark matter annihilating into two photons from the total line emission flux for an isothermal (dashed) and NFW dark matter profile (solid).}
	\label{fig:Esigmav}
\end{figure*}

The resulting constraint is plotted in \figref{Esigmav} (right). The integral over density squared is very sensitive to the choice of dark matter profile with a span of three orders of magnitude between the extremes in \tabref{mfov}. For the weakest constraints $\left<\sigma v \right> \gtrsim 10^{-34} \cm^{3}\sec^{-1}$ \cite[e.~g.][]{Cline:2014,Finkbeiner:2014} is allowed for all energies (masses), but for any other density profile, it is ruled out. The situation is similar for directly annihilating dark matter \citep[e.g.][]{Dudas:2014,Frandsen:2014,Baek:2014} since they are all designed to produce the stacked cluster signal.

\section{Constraints on sterile neutrinos} \label{sec:sterile}
\begin{figure*}
	\centering
	\includegraphics[width=0.99\textwidth]{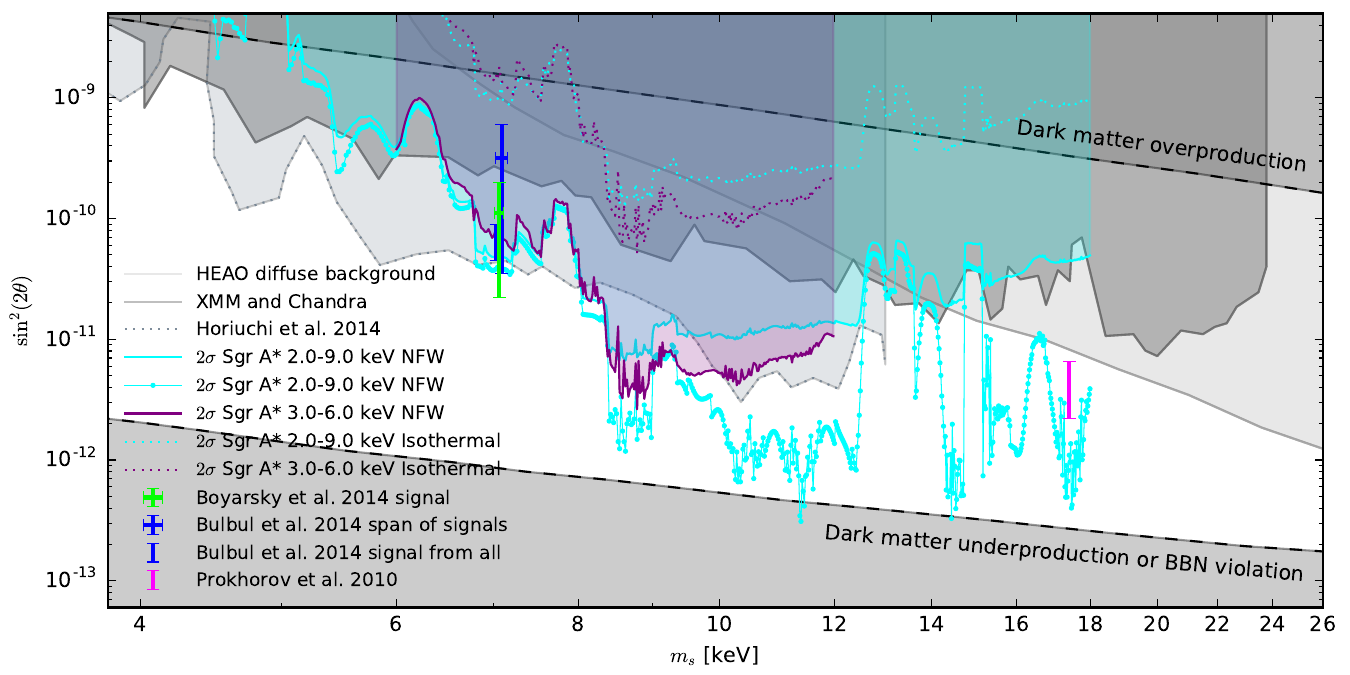}
	\caption{Mass-mixing angle constraints on sterile neutrino like dark matter candidates. Above/below the dashed black lines the sterile neutrinos will be over/under produced relative to the observed dark matter density \citep{Laine:2008a,Canetti:2013}. The grey shaded regions are X-ray exclusion lines from \XMM{} and \Chandra{} observations \citep[][with the first rescaled by a factor of two due to mass estimate uncertainties as recommended in \citet{Boyarsky:2009r}]{Boyarsky:2008b,Riemer-Sorensen:2009,Boyarsky:2006a} and the coloured regions is the parameter space ruled out by the present Sgr* analysis assuming an NFW profile for the Milky Way halo. The cyan lines are for the case where the entire 2-9\keV{} interval is modelled, and the purple lines are for the 3-6\keV{} interval. The dashed lines are for the isothermal profile where the constraints are weakened by an order of magnitude. The dotted grey line show the constraints from \citet{Horiuchi:2014} for which the model contains some lines of astrophysical origin. The green error bar is the signal from \citet{Boyarsky:2014}, and the blue error bars represent the signal from \citet{Bulbul:2014} with the small range indicating the best fit to the entire stack of spectra and the larger bars indicate the range of results from their different analyses. All the inferred signals are ruled out at 95\% confidence under the NFW assumption.}
	\label{fig:m_sin}
\end{figure*}

\citet{Bulbul:2014} and \citet{Boyarsky:2014} interpreted the detected line emission excess in the context of sterile neutrinos. In this section we introduce the sterile neutrino and discuss the Milky Way non-detection in this context.
The sterile neutrino is a strong particle candidate for the dark matter viable with or without Super Symmetry or Universal Extra Dimensions. With just three sterile neutrinos (gauge singlets), one can obtain the correct abundance of dark matter, a very simple explanation for the observed flavour oscillations and mass splittings of the active neutrinos, and a natural explanation for the baryon asymmetry \citep{Dodelson:1993, Shi:1998, Dolgov:2002, Abazajian:2001a, Asaka:2005a, Asaka:2005b,Canetti:2013}. The underlying particle model, called the $\nu$MSM, is described in detail in a number of papers \citep{Asaka:2005a, Asaka:2005b, Asaka:2006c, Shaposhnikov:2006c, Gorbunov:2007a, Laine:2008a, Shaposhnikov:2008a,Canetti:2012}, and recent reviews \citep{Boyarsky:2009r, Kusenko:2009, Drewes:2013}. Additionally, the sterile neutrino may have interesting effects on a range of astrophysical objects, e.~g. as an explanation for pulsar kick velocities, facilitating core collapse supernova explosions, affecting early star formation, reionization and structure formation, or assisting inflation \citep{Kusenko:1997, Hansen:2004, Fryer:2005, Hidaka:2006, Biermann:2006, Mapelli:2006, Shaposhnikov:2006c, Bezrukov:2007, Kusenko:2007, Petraki:2007, Petraki:2007b, Boyanovsky:2008, Gorbunov:2008}. The lightest of the three sterile neutrinos provides an attractive dark matter candidate. The two free parameters of mass, $m_s$, and mixing angle with the active neutrinos, $\sin^2(2\theta)$, are unconstrained from particle physics, but as seen in \figref{m_sin} various observations have already excluded large parts of this parameter space \citep{Laine:2008a,Canetti:2013,Boyarsky:2008b,Riemer-Sorensen:2009,Boyarsky:2006a,Horiuchi:2014}.

The mass is firmly bound from below through the phase space density of nearby dwarf galaxies. This so-called Tremaine-Gunn bound \citep{Tremaine:1979} gives a model independent boundary of roughly $m_s>0.4\keV$ \citep{Boyarsky:2008a}. The limit can be increased if the production method is known, and e.~g. for resonant production the boundary is approximately $1 \keV$ \citep{Boyarsky:2008a}.

Observations of small scale structure from the Lyman $\alpha$ forest can provide limits on the mass if the velocity distribution is known \citep{Hansen:2002, Viel:2005, Viel:2006, Seljak:2006}. The velocity distribution of the sterile neutrinos depends on their production mechanism in the early Universe, and while the originally proposed non-resonant production \citep{Dodelson:1993} is ruled out, in general masses above $2 \keV$ are allowed \citep{Boyarsky:2008e,Horiuchi:2014}. A plausible mechanism is via resonant production \citep{Shi:1998}, which requires a large initial lepton asymmetry \citep{Serpico:2005, Dolgov:2002}. However, the lepton asymmetry cannot be so large that it violates Big Bang Nucleosynthesis, and consequently there is a strong lower limit on the mixing angle to produce enough sterile neutrinos to account for the observed dark matter density \citep{Laine:2008a}. With a reasonable choice of parameter values, the resonant production provides sterile neutrinos with clustering properties similar to cold dark matter despite the $\keV$ mass \citep{Boyarsky:2008e}. Consequently dark matter simulations indicate an NFW profile for Milky Way sized halos, which may be flattened by the inclusion of baryons \citep{Navarro:1997,Kuhlen:2013,Dutton:2014}. The sterile neutrinos can also be produced at the electro-weak scale by decays of a gauge singlet Higgs boson \citep{Kusenko:2009}, or from their couplings to e.~g. the inflaton \citep{Shaposhnikov:2006c}.

The sterile neutrinos can decay via a one-loop diagram to an active neutrino and a photon. Since the two-body decay takes place almost at rest ($v/c\approx10^{-3}$ for galaxies), the decay line is very narrow and easily searched for in X-ray and soft gamma-ray observations. The fluxes are converted to constraints in the $m_s-\sin^2(2\theta)$ parameter space for sterile neutrinos of the Majorana type, assuming the sterile neutrinos to account for all of the dark matter in the observed field of view \citep{Abazajian:2001,Riemer-Sorensen:2006, Boyarsky:2007}:
\begin{eqnarray}\label{eq-massmixing}
&& \sin^2(2\theta) \leq \nonumber \\
&& 10^{18} \left(\frac{F_\mathrm{obs}}{\erg\cm^{-2}\sec{-1}}\right) \left( \frac{m_s}{\keV}\right) \left[ \frac{(M_\mathrm{fov}/\Ms)}{(D_L/\Mpc)^2} \right]
\end{eqnarray}
where $F_\mathrm{obs}$ is the observed flux limit, $M_\mathrm{fov}$ is the total dark matter mass within the field of view, and $D_{L,i}$ is the luminosity distance.

The $E_\gamma=0.3-12\, \mathrm{keV}$ range is constrained from various objects observed with the {\it Chandra} and {\it XMM-Newton} X-ray telescopes \citep{Boyarsky:2006a,Boyarsky:2006b,Riemer-Sorensen:2006,Abazajian:2006a,Boyarsky:2007,Riemer-Sorensen:2007,Boyarsky:2008a,Boyarsky:2008b,Riemer-Sorensen:2009,Loewenstein:2009,Loewenstein:2010,Loewenstein:2012}. In \figref{m_sin} we show a selection of the strongest robust constraints \citep{Boyarsky:2008b,Riemer-Sorensen:2009,Horiuchi:2014}. Some analyses have claimed stronger constraints, but were later found to be too optimistic \citep{Abazajian:2006a,Boyarsky:2006b,Abazajian:2001,Boyarsky:2008b,Watson:2006,Yuksel:2008,Boyarsky:2008a,Watson:2012}.
The higher energy range of $3-48\, \mathrm{keV}$ has been constrained from the diffuse X-ray background observed with HEAO \citep{Boyarsky:2006a} and the $3-80\, \mathrm{keV}$ with {\it NuSTAR} \citep{Riemer-Sorensen:2015}. The emission line constraints only depend on the amount and current properties of observed dark matter and are thus independent of the production mechanism.

The wide blue error bar represents the span of inferred mass and mixing angles from \citet{Bulbul:2014}, with the smaller blue error bar showing the \XMM{}-MOS signal for the stacked spectra of all clusters \citep{Bulbul:2014}. The green error bar is the \citet{Boyarsky:2014} line detection.

Another potential signal from sterile neutrinos has been suggested at $8.7\keV$ \citep{Prohorov:2010,Koyama:2007} from \Suzaku{} observations towards the Milky Way centre. A sterile neutrino origin of this signal would require a mixing angle of $\sin^2(2\theta)=4.4\pm2.2\times10^{-12}$, which is clearly ruled out by the present analysis (magenta error bars in \figref{m_sin}).

\section{The $3.5\keV$ line discussed}
Apart from the original detection in stacked galaxy cluster spectra from \XMM{} and individual clusters observed with \Chandra{} and \XMM{} \citep[$>3\sigma$ in various subsamples of the data,][]{Bulbul:2014,Boyarsky:2014}, \citet{Iakubovskyi:2015} detected the line at $2\sigma$ significance in a range of \XMM{} observations of individual clusters. \citet{Urban:2015} also detected the line in \Suzaku{} observations of the Perseus cluster, but not from other clusters. With detections from several different telescopes, an instrumental origin is unlikely for the Perseus signal. In order to pin down the origin of the line we need observations of several types of objects with different properties such as dark matter profiles and astrophysical background. However, searches in galaxies, dwarf galaxies and the Milky Way have been less conclusive and some even contradicting. Already the pre-\citet{Bulbul:2014} analysis of M31 \Chandra{} observations by \citet{Horiuchi:2014} was in conflict with the stacked cluster signal at the 95\% confidence level if assuming a decaying dark matter origin. The same is the case for stacked \XMM{} spectra of dwarf spheroidal galaxies \citep{Malyshev:2014} and stacked spectra of galaxies from both \XMM{} and \Chandra{} \citep[][rule out the signal at $7.8\sigma$]{Anderson:2014}, and one analysis of deep \XMM{} observations of the Draco dwarf galaxy \citep[][more than 99\% confidence]{Jeltema:2016}. However, \citet{Ruchayskiy:2016} do detect a line signal at $2.3\sigma$ from a part of the same \XMM{} observation, while the full data set is inconsistent with the signal predicted from Perseus. This perfectly illustrates the complexity of the on-going discussion. A number of searches do not find any evidence for line emission, but do not rule out a dark matter origin of the Perseus signal either e.g. \citet[][analysing \XMM{} observations of the Milky Way centre]{Jeltema:2014}, \citet[][\Suzaku{} observations of Perseus]{Tamura:2015}, \citet[][\Suzaku{} blank sky observations]{Sekiya:2015}, \citet[][sounding rocket microcalorimeter data]{Figueroa-Feliciano:2015}.

Decaying dark matter cannot explain all the observations directly (assuming it is single-type), but as already mentioned in \secref{general}, more complicated models can explain the line presence in galaxy clusters but not in smaller objects. For example, axion to photon conversion depends on the magnetic field strength and length scale and consequently provide different flux predictions for different objects \citep[e.g.][]{Cicoli:2014, Conlon:2014}. Also atomic transitions have been considered as the origin of the line with the K{\sc xviii} transition able to account for all of the galaxy cluster line excess \citep{Jeltema:2014,Phillips:2015} or high-n S{\sc xvi} transitions populated by charge transfer processes to the ground state \citep{Gu:2015}. Better energy resolution is needed to confirm or reject these scenarios.

Assuming an NFW profile for the Milky Way, the constraints presented here are similar to those of \citet{Horiuchi:2014} for sterile neutrino masses near $7\keV$, and up to an order of magnitude better than previous constraints at higher energies. If we instead assume a cored profile like the isothermal profile in \tabref{mfov}, the constraints weaken by up to a factor of 20. Consequently, the constraining power of the Sgr A* observations depends on the assumptions about the dark matter profile, and while we do not see any evidence for line emission from the galactic centre, we cannot rule out the existence for the flattest density profiles.

\citet{Boyarsky:2014b} analysed \XMM{} data of the galactic centre and found a $2.0-5.7\sigma$ evidence for line emission around $3.5\keV$. However, the line emission in \citet{Boyarsky:2014b} is only consistent with non-detection in blank sky data \citep{Boyarsky:2006a,Riemer-Sorensen:2006} for steeper--than--cored dark matter profiles such as Einasto or NFW, for which there is a tension with the results presented here. This discrepancy can neither confirm nor rule out a dark matter origin of the line emission but calls for further investigation with better resolution of the atomic lines and independent measurements of the element abundances and total mass profile near the galactic centre.

\section{Future improvements}
There are several factors playing a role when deciding where to look for decaying dar matter. First of all the signal strength is determined by the mass within the field of view and average distance. Remarkably, the line of sight integral for various astrophysical objects from cosmological background to clusters of galaxies to dwarfs satellites is very similar, and consequently the expected signal is very similar \citep{Boyarsky:2006b}, but the mass profile uncertainty has to be taken into account. The second factor is the expected background level from astrophysical sources. Here dwarf galaxies have an advantage \citep{Loewenstein:2009,Riemer-Sorensen:2009}. The real advantage of the Milky Way centre is its proximity and the large amount of data available, which can be directly compared to the "other" directions in the Milky Way via the blank sky data \citep{Boyarsky:2006b, Riemer-Sorensen:2006}. Unfortunately the prospects for improving the constraints using the galactic centre are not very optimistic since the main uncertainty lies with the inner slope of the dark matter halo profile. This can be partly mitigated by looking off-centre, which will also reduce the astrophysical background emission. Moving the field of view just $1\deg$ above or below the galactic plane reduces the differences between the integrated line of sights to $<10\%$ for the considered profiles. However, it also reduces the mass within the field of view by roughly an order of magnitude and consequently we would need to increase the observation time by two orders of magnitude to compensate statistically. At the moment less than 100~ks of \Chandra{} with an off-set of $\approx 1\deg$ spectra exist. Instead future improvements will come from increased spectral resolution of e.g. {\it Astro-H} and {\it Micro-X} microcalorimeters \citep{Mitsuda:2014,Figueroa-Feliciano:2015} , and mapping of the spatial morphology of the emission with existing instruments (\XMM, \Chandra, \Suzaku).

\section{Conclusions} \label{sec:conclusion}
The Milky Way data does not clearly show the emission line detected in stacked galaxy cluster spectra. The constraints on the allowed X-ray line emission flux are sensitive to the predicted amount of dark matter in the field of view. For an NFW profile the new constraints are up to an order of magnitude stronger than previous constraints in particular for photon energies above $4.0\keV$. For a cored isothermal profile the decaying dark matter constraints weaken by a factor of 20 and are thus weaker than previous constraints, but provide a cross-check based on a different object. Despite the improved constraints, the non-detection of line emission near Sgr A* is not inconsistent with the cluster line detection for the most conservative choice of dark matter profiles for the Milky Way. 

\begin{acknowledgements}
Thanks to Steen H. Hansen, Yassaman Farzan and Neal Weiner for inspiring discussions and to the anonymous referees for constructive comments. SRS would like to thank NORDITA for hosting the News in Neutrino Physics 2014 workshop during which most of the analysis was carried out. Parts of this research were conducted by the Australian Research Council Centre of Excellence for All-sky Astrophysics (CAASTRO), through project number CE110001020.
\end{acknowledgements}

\bibliographystyle{aa}
\bibliography{sterile}

\end{document}